# On the influence that the ground electrode diameter has in the propulsion efficiency of an asymmetric capacitor in nitrogen gas

**Alexandre A. Martins[1] and Mario J. Pinheiro[2]**
[1]Institute for Plasmas and Nuclear Fusion & Instituto Superior Técnico,
Av. Rovisco Pais, 1049-001 Lisboa, Portugal
[2]Department of Physics and Institute for Plasmas and Nuclear Fusion,
Instituto Superior Técnico, Av. Rovisco Pais, 1049-001 Lisboa, Portugal

In this work the propulsion force developed in an asymmetric capacitor will be calculated for three different diameters of the ground electrode. The used ion source is a small diameter wire, which generates a positive corona discharge in nitrogen gas directed to the ground electrode. By applying the fluid dynamic and electrostatic theories all hydrodynamic and electrostatic forces that act on the considered geometries will be computed in an attempt to provide a physical insight on the force mechanism that acts on the asymmetrical capacitors, and also to understand how to increase the efficiency of propulsion.

## I. INTRODUCTION

Our present work has the purpose to investigate the influence of the width of the ground electrode on the propulsion developed by an asymmetric capacitor which generates an electrohydrodynamic (EHD) flow through a corona discharge in nitrogen gas, at atmospheric pressure. We are going to study and compare the propulsion efficiencies of three different setups. The only variable between them is the width of the ground electrode (which as always a height of 0.02 m) used bellow a positive corona wire with a radius of 0.25 μm, and which is placed 0.03 m above the center of the ground electrode. The first structure to be studied is an asymmetric capacitor with the ground electrode having an ellipsoidal cross-section 0.02 m in height and 0.02 m in width with the forward section centered at (0 m, 0 m) and with the corona wire at (0 m, 0.03 m). In the second geometry to be studied the ground electrode has 0.04 m in width, and in the third capacitor the width is 0.06 m (Figure 1). In the ensuing discussion we are going to apply the known theory of EHD and electrostatics in order to provide a physical insight on the force mechanism that acts on the asymmetrical capacitors, and compare the developed forces in order to determine which is more efficient, and what influence the diameter of the ground electrode plays in the development of the propulsive force. These electrostatic actuators are very interesting because they generate a propulsion force which differs from the usual plasma actuators[1-3] in that they harness the electrostatic force to move the actuator itself.

## II. NUMERIC MODEL

In our numerical model we will consider electrostatic force interactions between the electrodes and the ion space cloud, the moment exchange between the mechanical setup and the induced opposite direction nitrogen flow (EHD flow), nitrogen pressure forces on the structure and viscous drag forces. EHD flow is the flow of neutral particles caused by the drifting of ions in an electric field. In our case these ions are generated by a positive high voltage corona discharge in the high curvature (the higher the radius of a sphere, the less is its curvature) portions of the electrodes. The corona wire has a uniform high curvature and therefore will generate ions uniformly.[4] On the contrary, the body electrode (ground) has a non uniform curvature having the lowest curvature facing the positive corona wire. The positively ionized gas molecules will travel from the corona wire ion source towards the collector (ground) colliding with neutral molecules in the process. These collisions will impart momentum to the neutral atoms that will move towards the collector as a result. The momentum gained by the neutral gas is exactly equal to the momentum gained by the positive ions accelerated through the electric field between the electrodes, and lost in inelastic collisions to the neutral molecules.

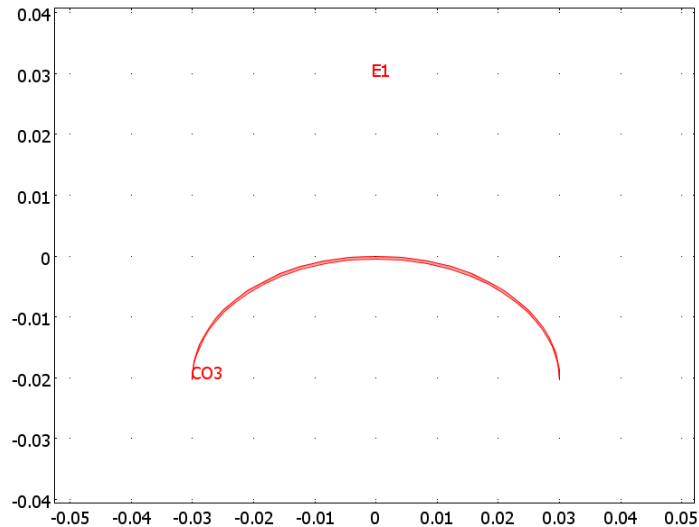

**Figure 1.** (Color online) Asymmetric capacitor 0.02 m height and 0.06 m width with the corona wire (E1) at (0 m, 0.03 m) and forward section of ground electrode (CO3) centered at (0 m, 0 m).

Corona discharges are non-equilibrium plasmas with an extremely low degree of ionization (roughly $10^{-8}$ %). There exist two zones with different properties, the ionization zone and the drift zone. The energy from the electric field is mainly absorbed by electrons in the ionization zone, immediately close to the corona electrode, and transmitted to the neutral gas molecules by inelastic collisions producing electron-positive ion pairs, where the net space charge density $\rho_q$ will remain approximately zero ($\rho_q = 0$). However, the local volume space charge, in the drift zone, will be positive for a positive corona in nitrogen (because it is electropositive and can only form positive ions, therefore $\rho_q = eZ_i n_i$ in the drift zone; $e$ is absolute value of the electron charge, $Z_i$ is the charge of the positive ionic species, $n_i$ is the positive ion density of $N_2^+$) because of the much higher mobility of electrons relative to the positive ions. We will only consider nitrogen ($N_2^+$) positive ions in the drift zone, because it is the dominating ion.[4] The following equations will be applied to the drift zone only.

The ionic mobility ($\mu_i$) is defined as the velocity $v$ attained by an ion moving through a gas under unit electric field E ($m^2$ $s^{-1}$ $V^{-1}$), i.e., it is the ratio of the ion drift velocity to the electric field strength:

$$\mu_i = \frac{v}{E}. \tag{1}$$

The mobility is usually a function of the reduced electric field $E/N$ and $T$, where $E$ is the field strength, $N$ is the Loschmidt constant (number of molecules $m^{-3}$ at s.t.p.), and $T$ is the temperature. The unit of $E/N$ is the Townsend (Td), 1 Td = $10^{-21}$ V $m^2$. Since we are applying 28000 V to the corona wire across a gap of 3 cm towards the ground electrode, the reduced electric field will be approximately 38 Td or 38 x $10^{-17}$ Vcm$^2$ (considering that the gas density N at 1 atm, with a gas temperature $T_g$ of 300 K is N=2.447 x $10^{19}$ cm$^{-3}$). According to Moseley,[5] the mobility $\mu_i$ of an ion is defined by:

$$\mu_i = \mu_{i0}(760/p)(T/273.16), \tag{2}$$

where $\mu_{i0}$ is the reduced mobility, $p$ is the gas pressure in Torr (1 atm = 760 Torr) and $T$ is the gas temperature in Kelvin. For our experimental condition of E/N = 38 Td, Moseley's measurements indicate a $\mu_{i0}$ of 1.83 cm$^2$/(Vs). Thus, at our operating temperature of 300 K, the mobility $\mu_i$ will be 2.01 cm$^2$/(Vs) or 2.01 x $10^{-4}$ m$^2$/(Vs). Since the reduced electric field is relatively low, the ion diffusion coefficient $D_i$ can be approximated by the Einstein relation:

$$D_i = \mu_i \left( \frac{k_B T}{e} \right), \tag{3}$$

where $k_B$ is the Boltzmann constant. This equation provides a diffusion coefficient of 5.19 x $10^{-6}$ m$^2$/s for our conditions.

The governing equations for EHD flow in an electrostatic fluid accelerator (EFA) are already known[6,7] and described next; these will be applied to the drift zone only. The electric field **E** is given by:

$$\mathbf{E} = -\nabla V. \tag{4}$$

Since $\nabla \cdot \mathbf{E} = \frac{\rho_q}{\varepsilon_0}$ (Gauss's law), the electric potential **V** is obtained by solving the Poisson equation:

$$\nabla^2 V = -\frac{\rho_q}{\varepsilon_0} = -\frac{e(Z_i n_i - n_e)}{\varepsilon_0}, \tag{5}$$

where $n_e$ is the negative electron density (not considered in the drift zone) and $\varepsilon_0$ is the permittivity of free space. The total volume ionic current density $\mathbf{J}_i$ created by the space charge drift is given by:

$$\mathbf{J}_i = \rho_q \mu_i \mathbf{E} + \rho_q \mathbf{u} - D_i \nabla \rho_q, \tag{6}$$

where $\mu_i$ is the mobility of ions in the nitrogen gas subject to an electric field, $u$ is the gas (nitrogen neutrals) velocity and $D_i$ is the ion diffusion coefficient. The current density satisfies the charge conservation (continuity) equation:

$$\frac{\partial \rho_q}{\partial t} + \nabla \cdot \mathbf{J}_i = 0. \tag{7}$$

But, since we are studying a DC problem, in steady state conditions we have:

$$\nabla \cdot \mathbf{J}_i = 0. \tag{8}$$

The hydrodynamic mass continuity equation for the nitrogen neutrals is given by:

$$\frac{\partial \rho_f}{\partial t} + \nabla \cdot (\rho_f \mathbf{u}) = 0. \qquad (9)$$

If the nitrogen fluid density $\rho_f$ is constant, like in incompressible fluids, then it reduces to:

$$\nabla \cdot \mathbf{u} = 0. \qquad (10)$$

In this case, the nitrogen is incompressible and it must satisfy the Navier-Stokes equation:

$$\rho_f \left( \frac{\partial \mathbf{u}}{\partial t} + (\mathbf{u} \cdot \nabla)\mathbf{u} \right) = -\nabla p + \mu \nabla^2 \mathbf{u} + \mathbf{f}. \qquad (11)$$

The term on the left is considered to be that of inertia, where the first term in brackets is the unsteady acceleration, the second term is the convective acceleration and $\rho_f$ is the density of the hydrodynamic fluid - nitrogen in our case. On the right, the first term is the pressure gradient, the second is the viscosity ($\mu$) force and the third is ascribed to any other external force $\mathbf{f}$ on the fluid. Since the discharge is DC, the electrical force density on the nitrogen ions that is transferred to the neutral gas is $\mathbf{f}^{EM} = \rho_q \mathbf{E} = -\rho_q \nabla V$. If we insert the current density definition (Equation (6)) into the current continuity (Equation (8)), we obtain the charge transport equation:

$$\nabla \cdot \mathbf{J}_i = \nabla \cdot (\rho_q \mu_i \mathbf{E} + \rho_q \mathbf{u} - D_i \nabla \rho_q) = 0. \qquad (12)$$

Since the fluid is incompressible ($\nabla \cdot \mathbf{u} = 0$) this reduces to:

$$\nabla \cdot (\rho_q \mu_i \mathbf{E} - D_i \nabla \rho_q) + \mathbf{u} \nabla \rho_q = 0. \qquad (13)$$

In our simulation we will consider all terms present in Equation (13), although it is known that the conduction term (first to the left) is preponderant over the other two (diffusion and convection), since generally the gas velocity is two orders of magnitude smaller than the velocity of ions. Usually, the expression for the current density (Equation (6)) is simplified as:

$$\mathbf{J}_i = \rho_q \mu_i \mathbf{E}, \qquad (14)$$

Then, if we insert Equation (14) into Equation (8), expand the divergence and use Equation (4) and Gauss's law we obtain the following (known) equation that describes the evolution of the charge density in the drift zone:

$$\nabla \rho_q \cdot \nabla V - \frac{\rho_q^2}{\varepsilon_0} = 0. \qquad (15)$$

In Table I we can see the values of the parameters used for the simulation. We will consider in our model that the ionization region has zero thickness, as suggested by Morrow.[8] The following equations will be applied to the ionization zone only. For the formulation of the proper boundary conditions for the external surface of the space charge density we will use the Kaptsov hypothesis[9] which states that below corona initiation the electric field and ionization radius will increase in direct proportion to the applied voltage, but will be maintained at a constant value after the corona is initiated.

In our case, a positive space charge $\rho_q$ is generated by the corona wire and drifts towards the ground electrode through the gap $G$ (drift zone) between both electrodes and is accelerated by the local electric field. When the radius of the corona wire is much smaller than $G$, then the ionization zone around the corona wire is uniform. In a positive corona, Peek's empirical formula[10-13] in air gives the electric field strength $E_p$ (V/m) at the surface of an ideally smooth cylindrical wire corona electrode of radius $r_c$:

$$E_p = E_0 \cdot \delta \cdot \varepsilon (1 + 0.308 / \sqrt{\delta \cdot r_c}). \tag{16}$$

Where $E_0 = 3.31 \cdot 10^6 V/m$ is the dielectric breakdown strength of air (we used the nitrogen breakdown strength which is 1.15 times higher than that for air[11]), $\delta$ is the relative atmospheric density factor, given by $\delta = 298p/T$, where $T$ is the gas temperature in Kelvin and $p$ is the gas pressure in atmospheres ($T$=300K and $p$=1atm in our model); $\varepsilon$ is the dimensionless surface roughness of the electrode ($\varepsilon = 1$ for a smooth surface) and $r_c$ is given in centimeters. At the boundary between the ionization and drifting zones the electric field strength is equal to $E_0$ according to the Kaptsov assumption. This formula (Peek's law) determines the threshold strength of the electric field to start the corona discharge at the corona wire. Surface charge density will then be calculated by specifying the applied electric potential $V$ and assuming the electric field $E_p$ at the surface of the corona wire. The assumption that the electric field strength at the wire is equal to $E_p$ is justified and discussed by Morrow.[8] Although $E_p$ remains constant after corona initiation, the space charge current $J_i$ will increase with the applied potential $V_c$ in order to keep the electric field at the surface of the corona electrode at the same Peek's value, leading to the increase of the surrounding space charge density and respective radial drift.

Atten, Adamiak and Atrazhev,[13] have compared Peek's empirical formula with other methods including the direct Townsend ionization criterion and despite some differences in the electric field, they concluded that the total corona current differs only slightly for small corona currents (below 6 kV). For voltages above 6 kV (corresponding to higher space currents) the difference is smaller than 10% in the worst case, according to them. For relatively low space charge density in DC coronas, the electric field $E(r)$ in the plasma (ionization zone) has the form:[4]

$$E(r) = \frac{E_p r_c}{r}, \tag{17}$$

where $r$ is the radial position from the center of the corona wire. Since the electric field $E_0$ establishes the frontier to the drift zone, using this formula we can calculate the radius of the ionization zone ($r_i$), which gives:

$$r_i = \frac{E_p r_c}{E_0} = r_c \cdot \delta \cdot \varepsilon (1 + 0.308 / \sqrt{\delta \cdot r_c}). \tag{18}$$

Since we have chosen in our simulation for $r_c$ to be 0.025 *mm*, then $r_i$ would be 0.074 *mm*. Now we can calculate the voltage ($V_i$) at the boundary of the ionization zone by integrating the electric field between $r_c$ and $r_i$:

$$V_i = V_c - E_p r_c \ln(E_p/E_0), \tag{19}$$

where $V_c$ is the voltage applied to the corona electrode and $r_c$ is in meters. This equation is valid only for the ionization zone. In our case it determines that if we apply 28000 Volts to the corona wire, then the voltage present at the boundary of the ionization zone becomes 26658.73 Volts.

For the drift zone, Poisson equation (Equation (5)) should be used together with the charge transport equation (Equation (13)) in order to obtain steady state field and charge density distributions. The values of the relevant parameters for the simulation are detailed in Table I. Three application modes of the COMSOL 3.5 Multiphysics software are used. The steady state incompressible Navier-Stokes mode is used to resolve the fluid dynamic equations. The electrostatics mode is used to resolve the electric potential distribution and the electrostatic forces to which the electrodes are subjected. The PDE (coefficient form) mode is used to resolve the charge transport equation (Equation (13)). The parameters used for the simulation are shown in Table I.

The solution domain was a square of 0.2 m on each side with the asymmetric capacitor at the center and containing typically 76698 elements (Figure 2). We have used a relative tolerance of $1 \times 10^{-6}$ to control the relative error the computed eigenvalues must agree at the evaluation points. In order to determine how the results vary with mesh resolution and to know the mesh size above which convergence is achieved we have performed repeated calculations at different mesh sizes for the asymmetric capacitor geometry with ground electrode dimensions of 0.02 m in height and 0.02 m in width (Figure 3). Table II shows the results obtained using as an example the values of the maximum wind velocity ($v_{max}$) and the total electrostatic force ($F_{eT}$) on the electrodes. Looking at this table we can see that the simulation results suffer several oscillations when the Mesh size is below 30000 elements. Above this value we have stability or convergence up to the third decimal place.

**TABLE I. Value of parameters used for the simulation.**

| Parameters | Value |
|---|---|
| Nitrogen density (T=300K, p=1atm), $\rho_N$ | 1.165 kg/m$^3$ |
| Dynamic viscosity of nitrogen (T=300K, p=1atm), $\mu_N$ | 1.775 x 10$^{-5}$ Ns/m$^2$ |
| Nitrogen relative dielectric permittivity, $\varepsilon_r$ | 1 |
| $N_2^+$ mobility coefficient, $\mu_i$ (for E/N = 38 Td) | 2.01 x 10$^{-4}$ m$^2$/(Vs) |
| $N_2^+$ diffusion coefficient, $D_i$ (for E/N = 38 Td) | 5.19 x 10$^{-6}$ m$^2$/s |
| Corona wire radius, $r_c$ | 25 μm |
| Facing ground electrode width | 0.02 / 0.04 / 0.06 m |
| Ground electrode height | 0.02 m |
| Nitrogen gap length | 0.03 m |
| Corona wire voltage, $V_c$ | 28000 V |
| Ground electrode voltage, $V_g$ | 0 V |

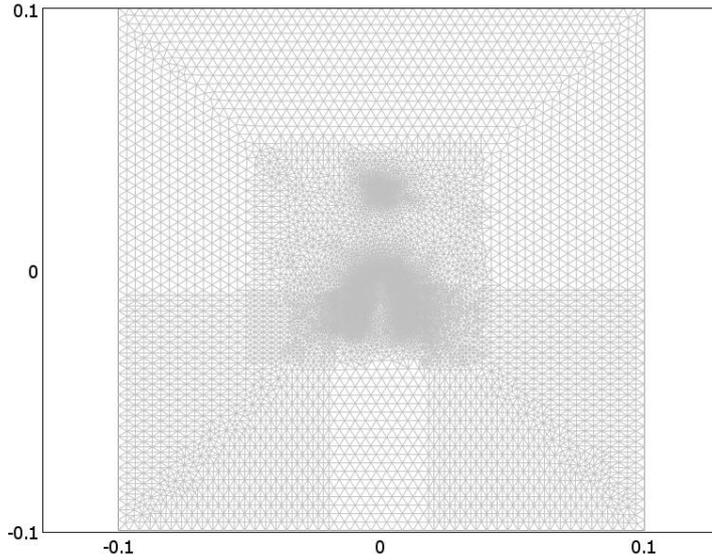

**Figure 2.** Typical mesh of the solution domain (0.2 m × 0.2 m) containing 76698 elements.

Dirichlet boundary conditions were used in the PDE (coefficient form) module, in the corona wire element, with an initial ion concentration of $8 \times 10^{-3}$ / $9 \times 10^{-3}$ / $4 \times 10^{-2}$ [C/m$^3$] respectively in the cases where the dimension of the ground electrode is 2 cm / 2 cm, 4 cm / 2 cm and 6 cm / 2 cm; and a zero ion concentration on the ground electrode and on the domain frontiers. In the Incompressible Navier-Stokes module, an open boundary condition was implemented on the domain frontiers, and a wall (no slip) condition was implemented on both electrodes. In the Electrostatics module a zero charge/symmetry boundary condition was implemented on the domain frontiers, a Ground potential on the ground electrode and an electric potential according to Equation (19) on the corona wire.

**Table II.** Variation of the maximum wind velocity ($v_{max}$) and the total electrostatic force ($F_{eT}$) on the electrodes depending on the Mesh size.

| Mesh Size | $v_{max}$ (m/s) | $F_{eT}$ (N/m) |
|---|---|---|
| **10258** | 3.510 | 0.245310 |
| **13584** | 3.490 | 0.243420 |
| **16012** | 3.479 | 0.243093 |
| **22078** | 3.476 | 0.243762 |
| **32450** | 3.484 | 0.241739 |
| **51738** | 3.484 | 0.241386 |
| **79098** | 3.484 | 0.241674 |

## III. NUMERICAL SIMULATION RESULTS

All the forces along the vertical (y axis) of the first geometry are presented in Table III. The results of the simulation show that the electrostatic forces $F_{ey}$ on the electrodes are the most relevant forces to consider, constituting 99.44% of the total force. The total hydrodynamic force $F_{HTy}$ is small because the pressure $F_{py}$ and viscosity $F_{vy}$ forces do not contribute in a relevant way in the present conditions. The total resultant force $F_{Ty}$ that acts on the capacitor is -0.243 N/m directed from the ground electrode to the wire.

**Table III. Forces along the y-axis of the asymmetric capacitor if the ground electrode is 2 cm wide and 2 cm high.**

|  | $F_{py}$ (N/m) | $F_{vy}$ (N/m) | $F_{HTy}$ (N/m) | $F_{ey}$ (N/m) | $F_{Ty}$ (N/m) |
|---|---|---|---|---|---|
| **Corona wire** | $-3.074\times10^{-5}$ | $5.425\times10^{-5}$ | $2.351\times10^{-5}$ | $-5.068\times10^{-4}$ | $-4.833\times10^{-4}$ |
| **Ground electrode** | $2.361\times10^{-4}$ | $1.096\times10^{-3}$ | $1.332\times10^{-3}$ | $2.422\times10^{-1}$ | $2.435\times10^{-1}$ |
| **Total force** | $2.053\times10^{-4}$ | $1.1502\times10^{-3}$ | $1.356\times10^{-3}$ | $2.417\times10^{-1}$ | $2.430\times10^{-1}$ |

The corona wire generates a positive charge cloud which accelerates through the nitrogen gap towards the facing ground electrode. The interaction between both electrodes and the positive charge cloud will accelerate the nitrogen positive ions towards the ground electrode and the ions will transmit their momentum to the neutral nitrogen particles by a collision process. Thus, the neutral nitrogen will move in the direction from the corona wire to the ground electrode, and the momentum transmitted to the ions and neutrals is equivalent to the electrostatic forces the ion cloud will induce on both electrodes. The nitrogen neutral wind generated by the collisions with the accelerated ion cloud is shown in Figure 3, where the top velocity achieved is 3.484 m/s. The corona wire and nitrogen positive space charge will induce opposite charges in the ground electrode, which will be subjected to a strong electrostatic force. In this way, the real force that makes the asymmetrical capacitor move is not a moment reaction to the induced draft but moment reaction to the positive ion cloud that causes the nitrogen movement. Total momentum is conserved always. The electrostatic force computation on both electrodes is shown in Figure 4.

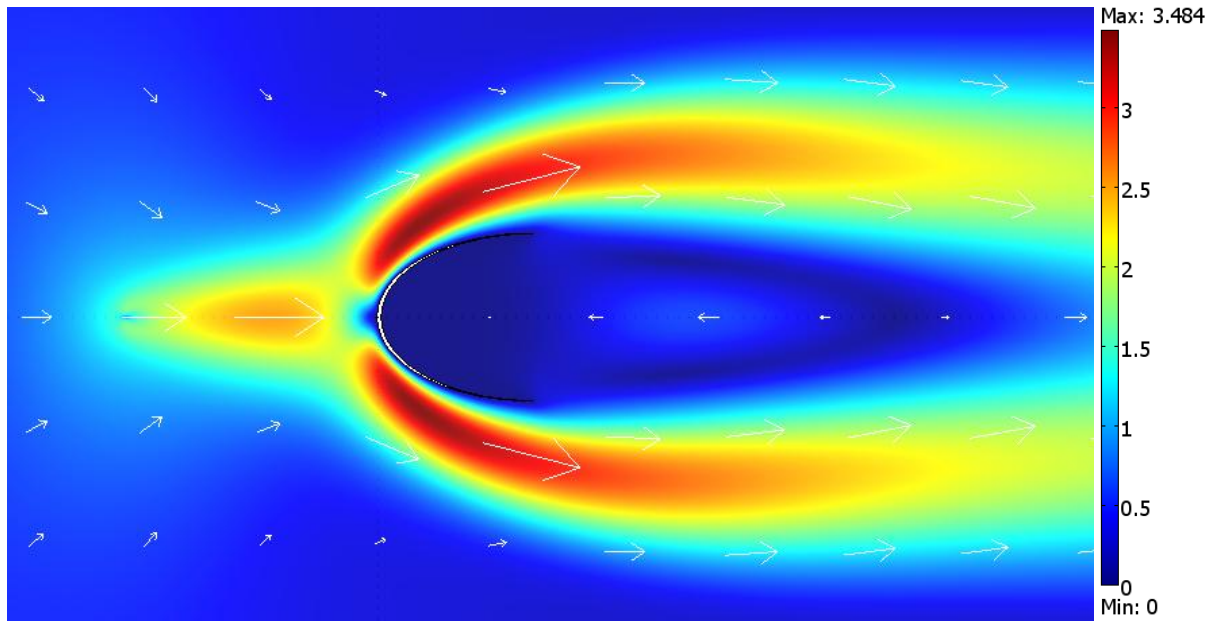

**Figure 3.** (Color online) Nitrogen velocity (when the ground electrode is 2 cm / 2 cm) as surface map with units in m/s with proportional vector arrows.

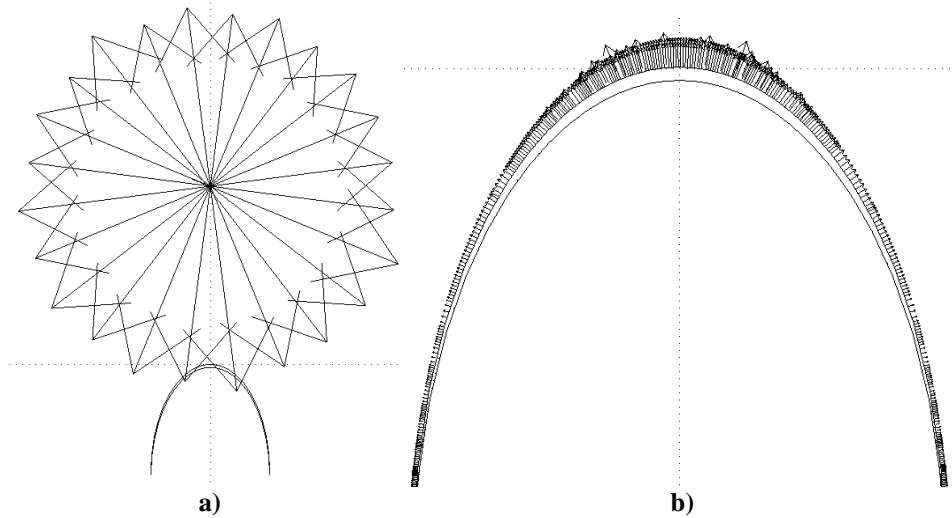

**Figure 4.** Electrostatic force (when the ground electrode is 2 cm / 2 cm) **a)** on the corona wire and **b)** on the ground electrode.

The electrostatic force on the corona wire is very strong but mostly symmetric (Figure 4.a)). Nevertheless, there is a slight asymmetry in the positive ion cloud distribution around the corona wire (Figure 5) in the direction of the ground electrode, because the positive ions are strongly pulled towards the ground electrode and this creates an asymmetrical distribution around the corona wire. Looking at Figure 5 and choosing a given density, for example 2 x $10^{-3}$ ($Cm^{-3}$), we can see that in the direction of the ground electrode this density value is farther away from the corona wire than at the opposite side of the wire. Because this asymmetry is small, the electrostatic force on the corona wire will also be small and directed towards the ground electrode, due to the positive ion distribution around it. The electrons that are provided to the neutral electrode (for it to remain neutral) are attracted to and neutralized in the front, were they suffer an electrostatic attraction towards the approaching positive ion cloud (Figure 4.b)), subjecting the ground electrode to a strong electrostatic force towards the corona wire. The electrostatic force on the wire is much stronger than that on the ground electrode, but since it is mostly symmetric around the wire the main electrostatic force will be on the ground electrode. The main thrust force on the electrodes is electrostatic, not hydrodynamic.

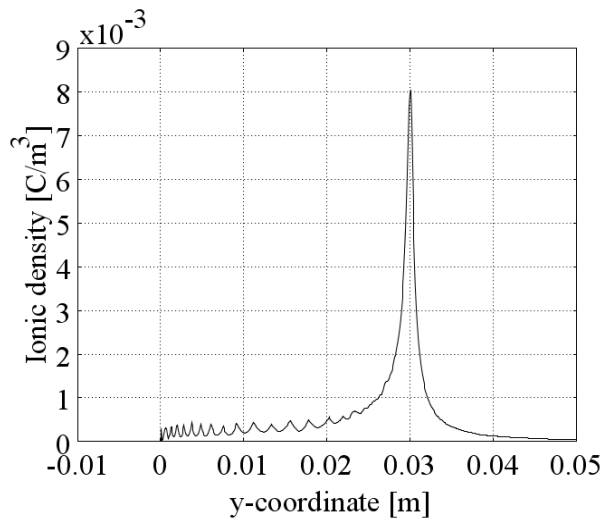

**Figure 5.** Ionic density distribution [$C/m^3$], for the case were the ground electrode is 2 cm / 2 cm, from (0 m, -0.01 m) to (0 m, 0.05 m).

For the asymmetrical capacitor with 4cm / 2cm, the highest nitrogen velocity achieved is 3.995 m/s. The total hydrodynamic and electrostatic forces for this case are presented in Table IV.

**Table IV. Forces along the y-axis of the asymmetric capacitor if the ground electrode is 4 cm wide and 2 cm high.**

|  | $F_{py}$ (N/m) | $F_{vy}$ (N/m) | $F_{HTy}$ (N/m) | $F_{ey}$ (N/m) | $F_{Ty}$ (N/m) |
|---|---|---|---|---|---|
| **Corona wire** | $-6.420\times10^{-5}$ | $7.395\times10^{-5}$ | $9.755\times10^{-6}$ | $-6.838\times10^{-4}$ | $-6.740\times10^{-4}$ |
| **Ground electrode** | $5.507\times10^{-3}$ | $1.129\times10^{-3}$ | $6.636\times10^{-3}$ | $2.877\times10^{-1}$ | $2.944\times10^{-1}$ |
| **Total force** | $5.443\times10^{-3}$ | $1.203\times10^{-3}$ | $6.646\times10^{-3}$ | $2.871\times10^{-1}$ | $2.937\times10^{-1}$ |

If the asymmetrical capacitor has 6 cm / 2cm, the highest nitrogen velocity achieved is 3.915 m/s, and the total hydrodynamic and electrostatic forces for this case are presented in Table V. We can again observe the slight asymmetry in the positive ion cloud distribution around the corona wire (Figure 6) in the direction of the ground electrode. If we analyze the lines of equal ionic density for this case (Figure 7) we can clearly see the neutralization of the positive ions on the ground electrode, and the ion charge asymmetry around the corona wire due to the strong electric attraction towards the ground electrode. The strongest electrostatic force on the ground electrode (Figure 8) matches the area were more positive ions are neutralized. Comparing the total forces for each case (Table VI) we can clearly see that the highest propulsive force occurs for the capacitor with the largest width (6 cm) and flatter surface, most certainly due to the larger available surface where the electrostatic force can act (and neutralize more positive ions), but also due to the more vertical electrostatic force vectors and higher ion concentration on the corona wire.

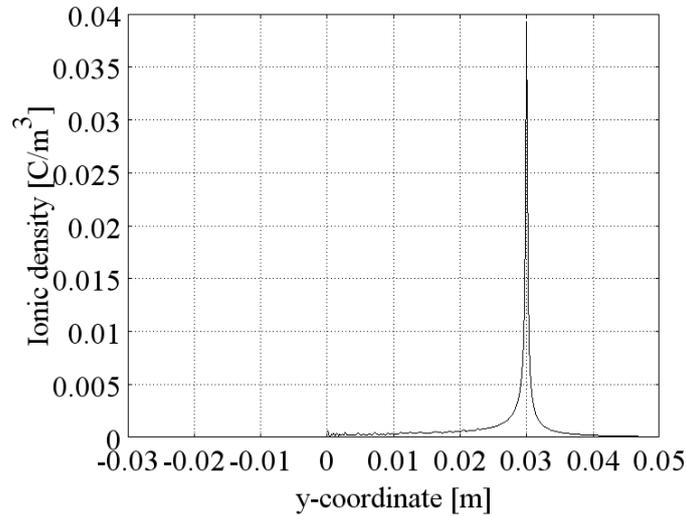

**Figure 6.** Ionic density distribution [C/m$^3$], for the case were the ground electrode is 6 cm / 2 cm, from (0 m, -0.03 m) to (0 m, 0.05 m).

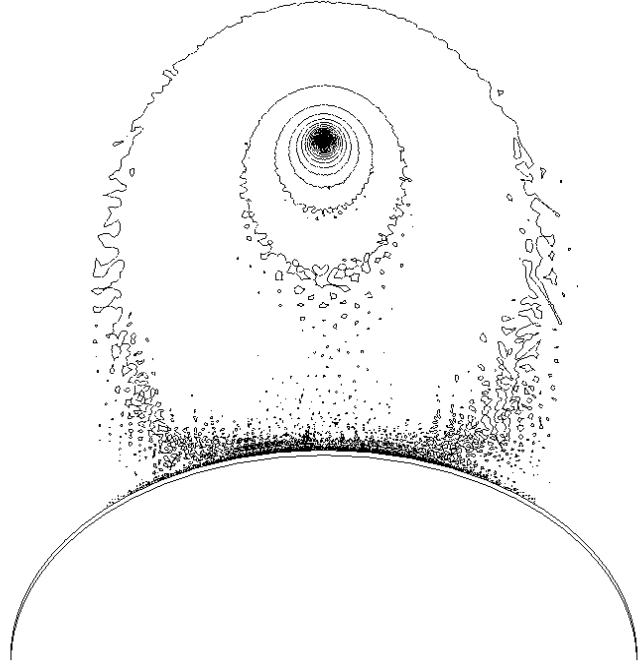

**Figure 7.** Lines of equal ionic density, when the ground electrode is 6 cm / 2 cm.

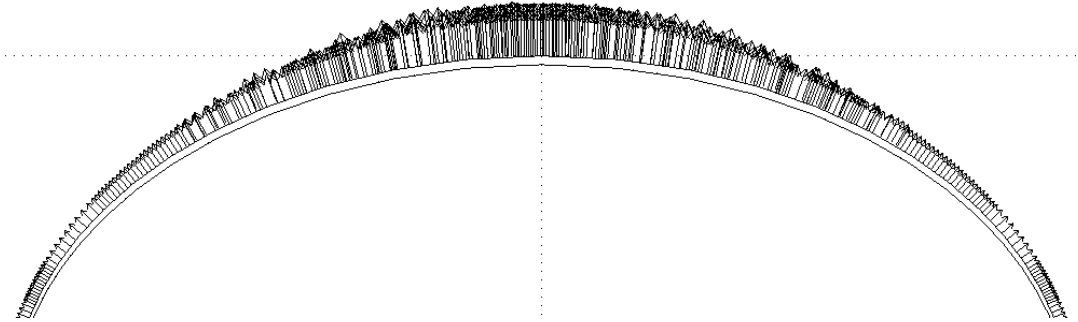

**Figure 8.** Electrostatic force on the ground electrode, for dimension 6 cm / 2 cm.

**Table V. Forces along the y-axis of the asymmetric capacitor if the ground electrode is 6 cm wide and 2 cm high.**

|  | $F_{py}$ (N/m) | $F_{vy}$ (N/m) | $F_{HTy}$ (N/m) | $F_{ey}$ (N/m) | $F_{Ty}$ (N/m) |
|---|---|---|---|---|---|
| **Corona wire** | $-5.893 \times 10^{-5}$ | $7.727 \times 10^{-5}$ | $1.835 \times 10^{-5}$ | $-1.009 \times 10^{-4}$ | $-8.259 \times 10^{-5}$ |
| **Ground electrode** | $5.979 \times 10^{-3}$ | $8.035 \times 10^{-4}$ | $6.782 \times 10^{-3}$ | $3.203 \times 10^{-1}$ | $3.271 \times 10^{-1}$ |
| **Total force** | $5.920 \times 10^{-3}$ | $8.808 \times 10^{-4}$ | $6.801 \times 10^{-3}$ | $3.202 \times 10^{-1}$ | $3.270 \times 10^{-1}$ |

**Table VI. Comparison of the total force on the considered asymmetric capacitors.**

|  | $F_{Ty}$ (N/m) | % $F_{ey}$ |
|---|---|---|
| **2 cm / 2 cm** | 0.243 | 99.44 |
| **4 cm / 2 cm** | 0.294 | 97.74 |
| **6 cm / 2 cm** | 0.327 | 97.92 |

## IV. CONCLUSION

The physical origin of the force that acts on an asymmetrical capacitor is electrostatic, with a component always superior to 97.7% on all cases, mainly concentrated on the ground electrode. The electrostatic force vectors on the wire are much stronger than that on the ground electrode, but since it is mostly symmetric around the wire the main electrostatic force will be on the ground electrode (due to the charge separation on the ground electrode - which remains neutral in total charge - from the corona wire and positive volume ion charge that exists above the ground and which attracts the metal mobile electrons upwards, inducing a charge separation on the ground electrode), which also has a much larger surface on which the force can act. The generated ion wind is a reaction to the electrostatic thrust mechanism and not the cause of the thrusting force as it is usually conceived. The main thrust force on the electrodes is electrostatic, not hydrodynamic. The mechanical momentum transferred to the nitrogen by the accelerated positive ions is balanced by the mechanical momentum of the physical structure of the capacitor. In simple terms, the sum of the interaction electrostatic force between the positive space charge ($\mathbf{F_{EI}}$, electrostatic force on the ions) and the capacitor electrodes ($\mathbf{F_{CS}}$, electrostatic force on the capacitor structure) is perfectly balanced:

$$\sum (\mathbf{F_{EI}} + \mathbf{F_{CS}}) = 0. \tag{20}$$

And since the positive ions transmit their momentum to the neutral nitrogen in collisions, then we will also have the mechanical momentum conservation:

$$\sum (\mathbf{p_{Nitrogen}} + \mathbf{p_{CS}}) = 0. \tag{21}$$

Where $\mathbf{p_{Nitrogen}}$ is the mechanical momentum that nitrogen acquires from the moving positive ions, and $\mathbf{p_{CS}}$ is the momentum of the capacitor structure. This last equation reflects the generally used ion wind argument, which is correct. However, in order to really understand the physical origin of the force on the capacitor structure we have to use Eq. (20). Therefore, we have the usual electrostatic forces that the capacitor electrodes impart on the gas ions which transfer momentum to the neutral gas in collisions forming the ion wind, but we also have the reciprocal electrostatic forces that the gas ions exert on the capacitor electrodes which are usually overlooked and forgotten, but are responsible for the capacitor motion.

The most efficient configuration occurs for the capacitor with the largest width (6 cm), probably due to the larger and flatter available surface were the electrostatic force vectors are more vertical, but also due to the neutralization of more positive ions on the larger surface. If we analyze the variation of the highest ion concentration on the corona wire between the different cases we can see that they are $8 \times 10^{-3} / 9 \times 10^{-3} / 4 \times 10^{-2}$ [C/m$^3$] when the width of the ground electrode is 2 cm, 4 cm and 6 cm respectively. Therefore, the higher the width of the ground electrode, the higher the positive ion concentration at the corona wire, and this is certainly also an important reason for the higher electrostatic force observed when the ground is 6 cm in width. The calculated forces on the considered setups are considered to be the maximum forces they could develop since the used voltage on the corona wire is near the spark limit for the used ground to wire distance (and for the wire radius of 25 μm) according to experimental measurements.[14] In the future we have to optimize the value of the electrostatic force and not of the aerodynamic force on the structure of the capacitor, in order to improve the force developed by this effect.

**ACKNOWLEDGEMENTS**

The authors gratefully thank to Mário Lino da Silva for the permission to use his computer with 32 gigabytes of RAM and two quad-core processors, without which this work would not have been possible. We also acknowledge partial financial support by the Reitoria da Universidade Técnica de Lisboa. We would like to thank important financial support to one of the authors, AAM, in the form of a PhD Scholarship from FCT (Fundação para a Ciência e a Tecnologia).